\documentclass[11pt,fleqn]{article}

\usepackage{amsmath,amsfonts,amssymb}
\usepackage{latexsym}
\usepackage{amsmath} 
\usepackage{amssymb} 
\usepackage{bm}
\usepackage{graphicx}
\usepackage{subfigure}

\newcommand{\twopartdef}[4]
{
	\left\{
		\begin{array}{ll}
			#1 & \mbox{if } #2 \\
			#3 & \mbox{if } #4
		\end{array}
	\right.
}

\newcommand{\hide}[1]{}

\newcommand{\n}{\nonumber\\}

\begin{document} 
%%%%%%%%%%%%%%%%%%%%%%%%%%%%%%%%%%%%%%%%%%%%%%%%%%%%%%%%%%%%%%%%%%%%%%%%%%%%%%%%%%%%%%%%%%%%
%%%%%%%%%%%%%%%%%%%%%%%%%%%%%%%%%%%%%%%%%%%%%%%%%%%%%%%%%%%%%%%%%%%%%%%%%%%%%%%%%%%%%%%%%%%%
%%%%%%%%%%%%%%%%%%%%%%%%%%%%%%%%%%%%%%%%%%%%%%%%%%%%%%%%%%%%%%%%%%%%%%%%%%%%%%%%%%%%%%%%%%%%
%%%%%%%%%%%%%%%%%%%%%%%%%%%%%%%%%%%%%%%%%%%%%%%%%%%%%%%%%%%%%%%%%%%%%%%%%%%%%%%%%%%%%%%%%%%%

\title {Initial condition from the action principle and its application to cosmology and to false vacuum bubbles}

\author{Eduardo Guendelman , Roee Steiner \\ Physics Department, Ben-Gurion University of the Negev, Beer-Sheva 84105, Israel}
\maketitle
\abstract{We study models where the gauge coupling constants, masses, etc are functions of some conserved charge in the universe. We first consider the standard Dirac action, but where the mass and the electromagnetic coupling constant are a function of the charge in the universe and afterwards extend this scalar fields. For Dirac field in the flat space formulation, the formalism is not manifestly Lorentz invariant, however Lorentz invariance can be restored by performing a phase transformation of the Dirac field. In the case where scalar field are considered, there is the new feature that an initial condition for the scalar field is derived from the action. In the case of the Higgs field, the initial condition require, that the universe be at the false vacuum state at a certain time slice, which is quite important for inflation scenarios. Also false vacuum branes will be studied in a similar approach. We discuss also the use of "spoiling terms", that violate gauge invariance to introduce these initial condition.}
%%%%%%%%%%%%%%%%%%%%%%%%%%%%%%%%%%%%%%%%%%%%%%%%%%%%%%%%%%%%%%%%%%%%%%%%%%%%%%%%%%%%%%%%%%%%%%%%%%%%%%%%%%%%
\section{Introduction}
Landau said " The future physical theory should contain not only the basic
equations but also the initial conditions for them " \cite{Landau}.
In physics we deal with equation of motion that are obtained by varying the action, here the question of the initial condition or boundary condition are normally separated from the equation of motion, and by giving them both we can solve the physical problem (like in many differential equation problems where the solution is determined by the initial condition).
Knowing just the equation of motion or just the initial conditions does not give the solution of the problem.
From this point we are motivated to construct a model where initial conditions can be found from the fundamental rules of physics, without the need to assume them, they will be derived.
Also we want to check whether the new model is consistent with causality and other requirements.\\
One of the examples of a system where the initial conditions are indirectly known, and the question is why should the initial condition be like that is the inflaton model. 
Today there are many models for inflation, the models are defined by the kind of inflation potential. 
The question is why the initial field should have specific initial conditions.
The problem in the inflation initial condition is that there is not known proven way to start the universe from a false vacuum state with vacuum energy density higher than the present universe needed for inflation. In fact it appears counter intuitive not to start in the lower energy state. One idea, the "eternal inflation" that one may think solve the problem, in fact does not solve the problem. Guth et.al wrote in their paper \cite{motivation guth} "\textit{Thus inflationary models require physics other than inflation to describe the past boundary of the inflating region of space time}". In their article it was proven that in the past of the eternal inflationary model there must be a singularity.\\
Also we are motivated to consider another direction in the research and study a model where the boundary conditions can follow from the action, this kind of approach can be used in a model where space like boundary condition of a system are fixed without any additional assumption, therefore fixing false vacuum boundary conditions on a brane. 
There are some equations in mathematical physics that constrain the possible initial condition that one can give. For example in electrodynamics , the equation $ \nabla \cdot E= 4\pi \rho $ is a time independent equation for $ E $ and $ \rho $, but tell us that we cannot give an initial value problem where $ \nabla \cdot E = 4\pi \rho $ is not satisfied. We want to deal in fact with a sort of constraint equations, but which do not impose a constraint every where, but only for a surface (time like or space like) therefore providing in fact initial or boundary condition, in the next section we review some ideas  on actions whose couplings depend on charges \cite{Mach like} which will be the basis  to achieve this, when charged scalar fields are introduced (following section).
We will see that generalizing the models where the gauge coupling constants, masses, etc are functions of some conserved charge in the universe may give such effect.\\\\
In a previous publication we considered the standard Dirac action, but where the mass and the electromagnetic coupling constant are a function of the charge in the universe and in this work we extend this scalar fields. This was motivated by the idea of obtaining a Mach like principle. For the Dirac field in the flat space formulation, the formalism is not manifestly Lorentz invariant, however Lorentz invariance can be restored by performing a phase transformation of the Dirac field. In the case where scalar field are considered, there is the new feature that an initial condition for the scalar field is derived from the action. In the case of the Higgs field inflation \cite{no minimal}, the initial condition require, that the universe be at the false vacuum state at a certain time slice, which is quite important for inflation scenarios. False vacuum branes will be studied in a similar approach.
We discuss also the use of "spoiling terms", that violate gauge invariant to introduce these initial condition. 

\section{The electromagnetic coupling constant as a function of the charge in the Dirac field}\label{per:charge}
We begin by considering the action for the Dirac equation
\begin{equation}\label{eq:simple dirac action}
S=\int d^{4}x\, \bar{\psi}(\frac{i}{2}\gamma^{\mu}\stackrel{\leftrightarrow}{\partial}_{\mu}-eA_{\mu}\gamma^{\mu}-m)\psi
\end{equation}
  where $ \bar{\psi}=\psi^{\dagger}\gamma^0 $.However here we take the coupling constant $ e $ to be proportional to the total charge (It can be generalized and we can also consider an arbitrary function of the total charge\cite{Mach like}).

\begin{equation}\label{eq:def e}
e=\lambda_{e}\int\psi^{\dagger}(\vec{y},y^{0}=t_{0})\psi(\vec{y},y^{0}=t_{0})\, d^{3}y=\lambda_{e}\int \rho(\vec{y},y^{0}=t_{0})\, d^{3}y
\end{equation}

and we will show that physics does not depend on the time slice $ y^{0}=t_{0} $

So after the new definition of "e" the  action will be:

\begin{eqnarray}
S=\int d^{4}x\, \bar{\psi}(x)(\frac{i}{2}\gamma^{\mu}\stackrel{\leftrightarrow}{\partial}_{\mu}-m)\psi(x)\nonumber\\-\lambda_{e}(\int d^{3}y\, \bar{\psi}(\vec{y},y^{0}=t_{0})\gamma^{0}\psi(\vec{y},y^{0}=t_{0}))(\int d^{4}x\, \bar{\psi}(x)A_{\mu}\gamma^{\mu}\psi(x))
\end{eqnarray}

we can express the three dimensional integral as a four dimensional integral

\begin{equation}\label{eq:trick}
\int d^{3}y\, \bar{\psi}(\vec{y},y^{0}=t_{0})\gamma^{0}\psi(\vec{y},y^{0}=t_{0})= \int d^{4}y\, \bar{\psi}(y)\gamma^{0}\psi(y)\delta(y^{0}-t_{0}) 
\end{equation}

so finally the action will be

\begin{equation}
S=\int d^{4}x\, \bar{\psi}(x)(\frac{i}{2}\gamma^{\mu}\stackrel{\leftrightarrow}{\partial}_{\mu}-m)\psi(x)-\lambda_{e}(\int d^{4}x\, \bar{\psi}(x)A_{\mu}\gamma^{\mu}\psi(x))(\int d^{4}y\, \bar{\psi}(y)\gamma^{0}\psi(y)\delta(y^{0}-t_{0}))
\end{equation}

if we consider the fact that $ \frac{\delta\bar{\psi}_{a}(x)}{\delta\bar{\psi}_{b}(z)}=\delta^{4}(x-z)\,\delta_{ab} $ and $ \frac{\delta\psi(x)}{\delta\bar{\psi}(z)}=0 $ we get the equation of motion

\begin{eqnarray}
\frac {\delta S}{\delta\bar{\psi}(z)}=0=\int{\delta^{4}(x-z)(i\gamma^{\mu}\partial_{\mu}-m)\psi (x)}\,d^{4}x\,\nonumber\\-\lambda_{e}(\int d^{4}x\, \delta^{4}(x-z)A_{\mu}\gamma^{\mu}\psi(x))(\int d^{4}y\, \bar{\psi}(y)\gamma^{0}\psi(y)\delta(y^{0}-t_{0}))\nonumber\\-\lambda_{e}(\int d^{4}x\, \bar{\psi}(x)A_{\mu}\gamma^{\mu}\psi(x))(\int d^{4}y\, \delta^{4}(y-z)\gamma^{0}\psi(y)\delta(y^{0}-t_{0}))
\end{eqnarray}

so to accomplish our goal we need just to perform the integrations in the last equation, and then the expression will simplified to

\begin{eqnarray}
\frac {\delta S}{\delta\bar{\psi}(z)}=(i\gamma^{\mu}\partial_{\mu}-m)\psi(z)-\lambda_{e}(\int{\bar{\psi}(y)\gamma^{0}\psi(y)\delta(y^{0}-t_{0})}\,d^{4}y)A_{\mu}\gamma^{\mu}\psi(z)&\nonumber\\-\lambda_{e}(\int{\bar{\psi}(x)A_{\mu}\gamma^{\mu}\psi(x)}\,d^{4}x)\gamma^{0}\psi(z)\delta(z^{0}-t_{0})
\end{eqnarray}

which can be simplified more by the use of new definition  $ b_{e}=\lambda_{e}(\int{\bar{\psi}(x)A_{\mu}\gamma^{\mu}\psi(x)}\,d^{4}x) $ which is a constant, and by the definition in equation (\ref{eq:def e}) 

\begin{equation}\label{eq:motion e}
\frac {\delta S}{\delta\bar{\psi}(z)}=[i\gamma^{\mu}\partial_{\mu}-m-eA_{\mu}\gamma^{\mu}-b_{e}\gamma^{0}\delta(z^{0}-t_{0})]\psi(z)=0
\end{equation}

so we can see that the last term in the equation of motion (\ref{eq:motion e}) contains $ A^{GF}_{\mu}\gamma^{\mu} $ where $ A^{GF}_{\mu}=\partial_{\mu}\Lambda $ and $ \Lambda=b_{e}\theta(z^{0}-t_{0}) $ is a pure gauge field. so the solution of this equation is
\begin{equation}\label{basic solution}
\psi=e^{-ib_{e}\theta(z^{0}-t_{0})}\psi_{D} 
\end{equation}

  where $ \psi_{D} $ is the solution of the equation
  \begin{equation}
 [i\gamma^{\mu}\partial_{\mu}-m-eA_{\mu}\gamma^{\mu}]\psi_{D}=0 
  \end{equation}
  
    from which it follows that $ j^{\mu}=\bar{\psi}_{D}\gamma^{\mu}\psi_{D}=\bar{\psi}\gamma^{\mu}\psi $ satisfies the local conservation law $ \partial_{\mu}j^{\mu}=0 $ and therefore we obtain that $ Q=\int{d^{3}x\, j^{0}} $ is conserved, so it does not depend on the time slice, furthermore it also follows that it is a scalar. For more examples see in referance \cite{Mach like}
%%%%%%%%%%%%%%%%%%%%%%%%%%%%%%%%%%%%%%%%%%%%%%%%%%%%%%%%%%%%%%%%%%%%%%%%%%%%%%%%%%%%%%%%%%%%%%%%%%%%%%%%%%%%

\section{Action which incorporates initials conditions}

As we will see now that type of actions considered in the previous section, when generalizing them to include charged scalar fields can provide some initials condition for the scalar field.
Those actions can be produced by taking the coupling constants as a function of a conserved charge.
If we use this development we can have the initial vacuum state for the universe in the inflationary model, so this initial condition will give us the initial condition for the universe corresponding to being initially at the false vacuum.
Following there are some examples of actions that can produce initial conditions.

\subsection{Initial condition from action with general potentials depending on charge}
We will begin with the action of Klein Gordon equation (with the metric $ diag(-1,1,1,1) $):
\begin{eqnarray}\label{general potential: action 2}
&S=\int{d^{4}x\sqrt{-g}\,[(\partial^{\mu}\phi^{*} +i\frac{g'}{2} A^{\mu}\phi^{*})(\partial_{\mu}\phi -i\frac{g'}{2} A_{\mu}\phi)-V(\phi,\phi^{*},Q)]}\nonumber\\&-\frac{1}{4}\int F^{\mu\nu}F_{\mu\nu}\sqrt{-g} d^{4}x - \frac{1}{16\pi G}\int{\sqrt{-g}R\,d^{4}x}=\nonumber\\&\int{d^{4}x\sqrt{-g}\,[(D\phi)^{*}(D\phi)-V(\phi,\phi^{*},Q)]}-\frac{1}{4}\int F^{\mu\nu}F_{\mu\nu}\sqrt{-g}  d^{4}x \n & - \frac{1}{16\pi G}\int{\sqrt{-g}R\,d^{4}x}
\end{eqnarray}
where the $ Q $ that appears in the potential $ V $ is given by:
\begin{equation}\label{eq: def m KG}
Q=\lambda\int{d^{3}y\,\sqrt{-g}[\phi^{*}i\stackrel{\leftrightarrow}{\partial^{0}}\phi + g'A^{0}\phi^{*}\phi]}\mid_{y^{0}=t^{0}}=\lambda\int{d^{4}y\,\sqrt{-g}[\phi^{*}i\stackrel{\leftrightarrow}{\partial^{0}}\phi + g'A^{0}\phi^{*}\phi]\delta(y^{0}-t_{0})}
\end{equation}
which is the total charge in the universe by the definition of Klein Gordon field.
So by variation we will get:
\begin{eqnarray}
&\delta S=\int{d^{4}x\,[-\delta\phi^{*}\,\partial_{\mu}(\sqrt{-g}\partial^{\mu}\phi) - i\delta\phi^{*} \frac{g'}{2}\partial_{\mu}(\sqrt{-g} A^{\mu}\phi) - i\sqrt{-g} \delta\phi^{*} \frac{g'}{2} A_{\mu}\partial^{\mu}\phi + (\frac{g'}{2})^{2}\sqrt{-g}\delta\phi^{*}A_{\mu}A^{\mu}\phi]}
\nonumber\\ &-\int{d^{4}x\,\sqrt{-g}\,\delta\phi^{*} \frac{\partial V}{\partial \phi^{*}}}
\nonumber\\&-\lambda(\int{d^{4}x\,\sqrt{-g}\,\frac{\partial V}{\partial Q}})\int{d^{4}y\,\delta(y^{0}-t_{0})[\delta\phi^{*}i\partial_{\nu}(\sqrt{-g}g^{0\nu}\phi)+\sqrt{-g}\delta\phi^{*}i\partial^{0}\phi+ g'\sqrt{-g}\delta\phi^{*}A^{0}\phi]}\nonumber\\
&-\lambda(\int{d^{4}x\,\sqrt{-g}\,\frac{\partial V}{\partial Q}})\int{d^{4}y\,\delta\phi^{*}i\phi \,\partial_{\nu}(g^{0\nu}\sqrt{-g}\delta(y^{0}-t_{0}))}=0
\end{eqnarray}
from this we get the equation of motion:

\begin{eqnarray}\label{eq:equation motion Klein Gordon 5435}
&-\partial_{\mu}(\sqrt{-g}g^{\mu\nu}\partial_{\nu}\phi) - i \frac{g'}{2}\partial_{\mu}(\sqrt{-g} A^{\mu}\phi) - i\sqrt{-g}\frac{g'}{2} A_{\mu}\partial^{\mu}\phi +\sqrt{-g} (\frac{g'}{2})^{2}A_{\mu}A^{\mu}\phi -\sqrt{-g}\frac{\partial V}{\partial \phi^{*}} 
\nonumber\\& -2i\sqrt{-g}\lambda(\int{d^{4}x\,\sqrt{-g}\,\frac{\partial V}{\partial Q}})\delta(y^{0}-t_{0})[\partial^{0}\phi -i\frac{g'}{2}A^{0}\phi]
-\lambda(\int{d^{4}x\,\sqrt{-g}\,\frac{\partial V}{\partial Q}})i\phi \,\partial^{0}(\sqrt{-g}\delta(y^{0}-t_{0}))=0
\end{eqnarray}

if we do the transformation 

\begin{equation}\label{eq:Klein A transform}
A^{0}\longrightarrow A^{0}+\frac{2i\lambda_{1}b}{g'}\delta(y^{0}-t_{0})
\end{equation}
and
\begin{equation}\label{eq:Klein psi transform}
\phi=e^{\lambda_{2} b\theta(y^{0}-t_{0})}\phi_{0}
\end{equation}
where $ b=i\lambda(\int{d^{4}x\,\sqrt{-g}\,\frac{\partial V}{\partial Q}}) $
\\we have that:

\begin{eqnarray}\label{eq:equation motion Klein Gordon 2}
&-\partial_{\mu}(\sqrt{-g}g^{\nu\mu}\partial_{\nu}\phi_{0}) - i \frac{g'}{2}\partial_{\mu}(\sqrt{-g}A^{\mu}\phi_{0}) - i\sqrt{-g}\frac{g'}{2} A_{\mu}\partial^{\mu}\phi_{0} + \sqrt{-g}(\frac{g'}{2})^{2}A_{\mu}A^{\mu}\phi_{0}-\sqrt{-g}\frac{\partial V}{\partial \phi^{*}} \nonumber\\&
-2b\sqrt{-g}\delta(y^{0}-t_{0})[(\lambda_{1}-\lambda_{2}+1)(\partial^{0}\phi_{0}-i\frac{g'}{2} A^{0}\phi_{0})\n & +0.5b\delta(y^{0}-t_{0})\phi_{0}(-\lambda^{2}_{2}+(2\lambda_{1}-\lambda^{2}_{1})+2(\lambda_{1}-1)\lambda_{2})]\nonumber\\&-b(\lambda_{1}-\lambda_{2}+1)\phi_{0}\,\partial^{0}(\sqrt{-g}\delta(y^{0}-t_{0})) = 0
\end{eqnarray}
if we require that the equation (\ref{eq:equation motion Klein Gordon 2}) will be like the ordinary Klein Gordon equation where there are no delta function appear, since those delta functions represent singular interactions, we need that:

\begin{eqnarray}
\lambda_{1}-\lambda_{2}+1=0\\
-\lambda^{2}_{2}+(2\lambda_{1}-\lambda^{2}_{1})+2(\lambda_{1}-1)\lambda_{2}=0
\end{eqnarray}

But there is no solution for $ \lambda_{1} $ and $ \lambda_{2} $ for those two equation.
If we will say that the covariant derivative is equal to zero $ \partial^{0}\phi_{0}-i\frac{g'}{2} A^{0}\phi_{0}=0 $ and $ \lambda_{1}-\lambda_{2}=2 $ then we still have problem with the term $ \partial^{0}\delta(y^{0}-t_{0}) $ in equation \ref{eq:equation motion Klein Gordon 2}. So we must to say that:
\begin{equation}
\boxed{\phi^{*}(t=0)\phi(t=0)=0}
\end{equation}
where $ \lambda_{1}-\lambda_{2}+1=0 $ which eliminates all the delta term in equation \ref{eq:equation motion Klein Gordon 2}.
which means that at $ t=0 $ we get the Higgs at false vacuum state\\   

\includegraphics[scale=0.4]{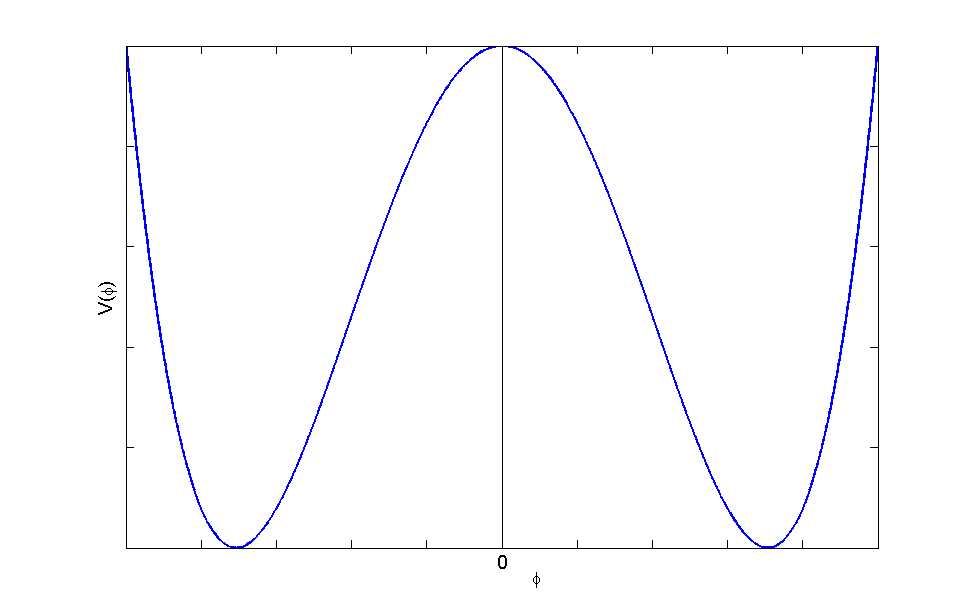}  

In the inflationary Higgs scenario \cite{no minimal} , the false vacuum initial state is important.

It is easy to see that, the same transformation work also for the case of variation of $ A_{\mu} $.
By doing variation on equation \ref{general potential: action 2} by $ A_{\mu} $ we get:
\begin{eqnarray}\label{general potantial :variation of action by A}
&\int d^{4}x\sqrt{-g} [\frac{g'}{2} \phi^{*}i\stackrel{\leftrightarrow}{\partial^{\mu}}\phi +2(\frac{g'}{2})^{2}A^{\mu}\phi^{*}\phi]\delta A_{\mu}
\\&+ g'\int{d^{4}x\sqrt{-g}\,\frac{\partial V(\phi,\phi^{*},Q)}{\partial Q}} \int d^{3}x\sqrt{-g}\phi^{*}\phi\delta^{\mu}_{0}\delta A_{\mu}+\int \partial_{\nu}F^{\nu\mu}\delta A_{\mu}\sqrt{-g} d^{4}x=0
\end{eqnarray}
so the equation of motion is:
\begin{equation}\label{eq mass:equation of motion F 1}
\partial_{\nu}F^{\nu 0}= \frac{g'}{2} \phi^{*}i\stackrel{\leftrightarrow}{\partial^{0}}\phi +2(\frac{g'}{2})^{2}A^{0}\phi^{*}\phi+ g'\int{(\frac{\partial V(\phi,\phi^{*},Q)}{\partial Q}\sqrt{-g}\,d^{4}x)} \phi^{*}\phi\delta(t)=j^{0}
\end{equation}

We can see also that the gauge charge, defined by the right hand side of \ref{eq mass:equation of motion F 1} , $ \bar{Q}=\lambda\int{j^{0}\sqrt{-g}d^{4}x} $ and the charge $ Q $ defined by \ref{eq: def m KG} differ by a $ \delta(0) $ term. 
If $ Q= \infty $ then $ V(\phi^{*},\phi,Q) $ for a non trivial function of $ Q $ will explode, so $ Q < \infty $. Also $ \bar{Q} < \infty $ because the electric charge must be finite, so if $ Q\mid_{t=0}=\bar{Q}\mid_{t=0}+c\delta(t)\mid_{t=0} $ and $ \delta(t)\mid_{t=0}=\infty $ then we must to say that $ c=0 $ so from that we get $ \phi^{*}\phi=0 $. We get back from a different point of view to the same condition. A specific example is that of a closed universe where we must have $ \bar{Q}=0 $ from this it means that $ Q\mid_{t=0}=c\delta(t)\mid_{t=0}=c\delta(0) $ which means that $ c=0 $, because $ Q $ enters in the potential and the potential must be well define, so $ c=0 $ , which means $ \phi^{*}(0,x)\phi(0,x)=0 $. Notice that other authors \cite{Models of Topology Change} have noticed that singular interaction can impose boundary condition.

\subsection{Boundary condition from action}
We now take another direction different from the question of the initial condition of inflation, and build up action that its boundary condition can be follow. To do so we use close technique as we done above but the solution is different and can be in use for other things. 
We will use the definition of the book of Anderson \cite{Anderson}, which take the points in sub-manifold:
\begin{equation}
x^{\mu}=\Phi^{\mu}(\lambda_{1},...,\lambda_{N})
\end{equation}
the element of area is:
\begin{equation}
d\tau^{\mu_{1},...,\mu_{N}}=\delta^{\mu_{1},...,\mu_{N}}_{\nu_{1},...,\nu_{N}}\frac{\partial\Phi^{\nu_{1}}}{\partial \lambda_{1}}....\frac{\partial\Phi^{\nu_{N}}}{\partial \lambda_{N}}d\lambda_{1}...d\lambda_{N}
\end{equation}
where:
\begin{eqnarray}
\delta^{\mu_{1},...,\mu_{N}}_{\nu_{1},...,\nu_{N}}=\begin{vmatrix} \delta^{\mu_{1}}_{\nu_{1}} & ... & \delta^{\mu_{N}}_{\nu_{1}} \\... &  &  \\ \delta^{\mu_{1}}_{\nu_{N}} &  & \delta^{\mu_{N}}_{\nu_{N}}  \end{vmatrix}
\end{eqnarray}
it can be also be written as:
\begin{equation}
d_{i}x^{\mu}=\frac{\partial \Phi^{\mu}}{\partial \lambda_{i}}d \lambda_{i}
\end{equation}
so the element of area is:
\begin{equation}
d\tau^{\mu_{1},...,\mu_{N}}=\delta^{\mu_{1},...,\mu_{N}}_{\nu_{1},...,\nu_{N}}d_{1}x^{\nu_{1}}...d_{N}x^{\nu_{N}}
\end{equation}
The dual element of the area element in $ N=4 $ dimension is:
\begin{equation}
d\sigma_{\mu}=\frac{1}{3!}\epsilon_{\mu\nu\rho\sigma}d\tau^{\nu\rho\sigma} \\\,\, ,\,\, d\sigma=\frac{1}{4!}\epsilon_{\mu\nu\rho\sigma}d\tau^{\mu\nu\rho\sigma}
\end{equation}
where $ \epsilon_{\mu\nu\rho\sigma} $ is Levi Civita tenzor where $ \epsilon^{\mu\nu\rho\sigma} $ is weight $ -1 $.
By the stokes theorem we have:
\begin{equation}
\oint{g^{\mu\nu}j_{\nu}\sqrt{-g} d\sigma_{\mu}}=\int{\partial_{\mu}(\sqrt{-g}j^{\mu})d\sigma}
\end{equation}
In our case $ j_{\mu}=\phi^{*}\stackrel{\leftrightarrow}{\partial}_{\mu}\phi-g'A_{\mu}\phi^{*}\phi $, where $ \partial_{\mu}(\sqrt{-g} j^{\mu})=0 $ is from the conservation law of scalar field.
so we have that:
\begin{equation}
\oint{j_{\nu}g^{\mu\nu}\sqrt{-g} d\sigma_{\mu}}=\int_{\mathcal{M}}{\partial_{\mu}(\sqrt{-g}j^{\mu}) d\sigma}=0
\end{equation}
So if we have close surface $ \Sigma=\Sigma_{1}+\Sigma_{2} $, where $ \mathcal{M} $ is the volume inside than we can have another conservation:
\begin{equation}
\oint_{\Sigma}{j_{\nu}g^{\mu\nu}\sqrt{-g} d\sigma_{\mu}}=\int_{\Sigma_{1}}{j_{\nu}g^{\mu\nu}\sqrt{-g} d\sigma_{\mu}}-\int_{\Sigma_{2}}{j_{\nu}g^{\mu\nu}\sqrt{-g} d\sigma_{\mu}}=0
\end{equation}
so:
\begin{equation}\label{bounders: theta def sigma}
\Theta \equiv \int_{\Sigma_{1}}{j_{\nu}g^{\mu\nu}\sqrt{-g} d\sigma_{\mu}}=\int_{\Sigma_{2}}{j_{\nu}g^{\mu\nu}\sqrt{-g} d\sigma_{\mu}}=const
\end{equation}
in the case $ d\sigma_{\mu} $ is space like, this represent the total amount of charge through the surface that entered over all times.\\
If we define theta function:
\begin{equation}
\theta(f(x))=\twopartdef {1}{f > 0}{0}{f < 0 }
\end{equation}
where $ f(x^{\mu})=0 $ on the surface $ \Sigma_{1} $
then we have:
\begin{equation}
\delta^{\mu}(f(x))=\partial^{\mu}\theta(f(x))
\end{equation}
So we can see equation \ref{bounders: theta def sigma} in anther way:
\begin{eqnarray}
\Theta = \int_{\mathcal{M}_{1}}{(j^{\mu}\delta_{\mu}(f(x)))\sqrt{-g} d\sigma}
\end{eqnarray}
where $ \int_{\mathcal{M}_{1}}{\delta_{\mu}(f(x))\sqrt{-g} d\sigma}=\int_{\Sigma_{1}}{\sqrt{-g} d\sigma_{\mu}} $.\\
Now going back to the action as we used before, where now the potential is $ V(\phi,\phi^{*},\Theta) $:
\begin{eqnarray}\label{boundery: action}
&S=\int{d\sigma\sqrt{-g}\,[(\partial_{\mu}\phi^{*} +i\frac{g'}{2} A_{\mu}\phi^{*})(\partial^{\mu}\phi -i\frac{g'}{2} A^{\mu}\phi)-V(\phi,\phi^{*},\Theta)]}\nonumber\\&-\frac{1}{4}\int F^{\mu\nu}F_{\mu\nu}\sqrt{-g} d\sigma  - \frac{1}{16\pi G}\int{\sqrt{-g}R\,d\sigma}=\nonumber\\&\int{d\sigma\sqrt{-g}\,[(D\phi)^{*}(D\phi)-V(\phi,\phi^{*},\Theta)]}-\frac{1}{4}\int F^{\mu\nu}F_{\mu\nu}\sqrt{-g}  d\sigma 
\n & - \frac{1}{16\pi G}\int{\sqrt{-g}R\,d\sigma}
\end{eqnarray}
from this by variation on $ \phi^{*} $ , we get the equation of motion:
\begin{eqnarray}\label{eq:equation motion Klein Gordon 543}
&-\partial_{\mu}(\sqrt{-g}g^{\mu\nu}\partial^{\nu}\phi) - i \frac{g'}{2}\partial_{\mu}(\sqrt{-g}A^{\mu}\phi) - i\sqrt{-g}\frac{g'}{2} A_{\mu}\partial^{\mu}\phi + \sqrt{-g}(\frac{g'}{2})^{2}A_{\mu}A^{\mu}\phi -\sqrt{-g}\frac{\partial V}{\partial \phi^{*}} 
\nonumber\\& -\lambda\sqrt{-g}(\int{d\sigma\,\sqrt{-g}\,\frac{\partial V}{\partial \theta}})\delta^{\mu}(f(x))[2i\partial_{\mu}\phi-g'A_{\mu}\phi]
\n &
-\lambda(\int{d\sigma\,\sqrt{-g}\,\frac{\partial V}{\partial \theta}})i\phi \,\partial_{\mu}(\sqrt{-g}\delta^{\mu}(f(x)))=0
\end{eqnarray}
if we do the transformation 
\begin{equation}\label{eq:Klein A transform mu}
A^{\mu}\longrightarrow A^{\mu}+\frac{2i\lambda_{1}b}{g'}\delta^{\mu}(f(x))
\end{equation}
and
\begin{equation}\label{eq:Klein psi transform mu}
\phi=e^{\lambda_{2} b\theta(f(x))}\phi_{0}
\end{equation}
where $ b=i\lambda(\int{d\sigma\,\sqrt{-g}\,\frac{\partial V}{\partial \theta}}) $
\\we have that:

\begin{eqnarray}\label{eq:equation motion Klein Gordon 2 mu}
&-\partial_{\mu}(\sqrt{-g}g^{\mu\nu}\partial_{\nu}\phi_{0}) - i \frac{g'}{2}\partial_{\mu}(\sqrt{-g}A^{\mu}\phi_{0}) - i\sqrt{-g}\frac{g'}{2} A_{\mu}\partial^{\mu}\phi_{0} + \sqrt{-g}(\frac{g'}{2})^{2}A_{\mu}A^{\mu}\phi_{0}-\sqrt{-g}\frac{\partial V}{\partial \phi^{*}} \nonumber\\&
-2b\sqrt{-g}\delta^{\mu}(f(x))[(\lambda_{2}-\lambda_{1}+1)\partial_{\mu}\phi_{0}-i(\lambda_{2}-\lambda_{1}+1)\frac{g'}{2} A_{\mu}\phi_{0}\n &+0.5b\delta_{\mu}(f(x))\phi_{0}(\lambda^{2}_{2}-2\lambda_{1}+\lambda^{2}_{1}+2(-\lambda_{1}+1)\lambda_{2})]\nonumber\\&-b(\lambda_{1}-\lambda_{2}+1)\phi_{0}\,\partial_{\mu}(\sqrt{-g}\delta^{\mu}(f(x))) = 0
\end{eqnarray}
if we need that equation (\ref{eq:equation motion Klein Gordon 2 mu}) will be like ordinary Klein Gordon equation we need that:

\begin{eqnarray}
\lambda_{2}-\lambda_{1}+1=0\\
\lambda^{2}_{2}-2\lambda_{1}+\lambda^{2}_{1}+2(-\lambda_{1}+1)\lambda_{2}=0
\end{eqnarray}
for which there is no solution, so we must conclude that $ \phi(x_{0})=0 $ when $ f(x_{0})=0 $
\\We can see also that variation by $ A^{\mu} $ on the action \ref{boundery: action} gives us:
\begin{eqnarray}\label{bounder :equation of motion F 1}
&\frac{1}{\sqrt{-g}}\partial_{\nu}(\sqrt{-g}F^{\nu \mu})= \frac{g'}{2} \phi^{*}i\stackrel{\leftrightarrow}{\partial^{\mu}}\phi +2(\frac{g'}{2})^{2}A^{\mu}\phi^{*}\phi \n &+ g'\int{(\frac{\partial V(\phi,\phi^{*},\Theta)}{\partial \Theta}\sqrt{-g'}\,d\sigma)} \phi^{*}\phi\delta^{\mu}(f(x))=j^{\mu}_{e}
\end{eqnarray}

%%%%%%%%%%%%%%%%%%%%%%%%%%%%%%%%%%%%%%%%%%%%%%%%%%%%%%%%%%%%%%%%%%%%%%%%%
\section{Boundary condition from spoiling terms}
Some "spoiling terms" that is terms that break gauge invariance have been shown in the end do not to contribute the functional integral \cite{spoil}. Here we will see that "spoiling terms" where non gauge invariant charge are introduced, have as a consequence that they induce boundary condition and these boundary condition imply the vanishing of the spoiling terms.
To see this we take the action
\begin{eqnarray}\label{general potential spoiling: action 2}
&S=\int{d^{4}x\sqrt{-g}\,[(\partial^{\mu}\phi^{*} +i\frac{g'}{2} A^{\mu}\phi^{*})(\partial_{\mu}\phi -i\frac{g'}{2} A_{\mu}\phi)-V(Q^{NGI},\phi,\phi^{*})]}\nonumber\\&-\frac{1}{4}\int F^{\mu\nu}F_{\mu\nu}\sqrt{-g} d^{4}x - \frac{1}{16\pi G}\int{\sqrt{-g}R\,d^{4}x}
\end{eqnarray}
where we introduce the non gauge invariant charge
\begin{equation}
Q^{NGI}= \int{d^{4}x\sqrt{-g}\delta(t-t_{0})[\phi^{*}i\stackrel{\leftrightarrow}{\partial^{0}}\phi + g'_{1}A^{0}\phi^{*}\phi]}
\end{equation}
where NGI= Non Gauge Invariant, and $ g'_{1} \neq g' $.
If we vary the action by a gauge transformation 
\begin{equation}
A_{\mu} \rightarrow \partial_{\mu}\Lambda +A_{\mu}\\
\phi \rightarrow \phi_{0}e^{ig'\Lambda}
\end{equation}
all the other term of the action can not be change but
\begin{equation}
Q^{NGI} \rightarrow Q^{NGI}+(g'-g'_{1})\int{d^{4}x\sqrt{-g}\delta(t-t_{0})\partial_{0}\Lambda (x) \phi^{*}\phi }
\end{equation}
So for all $ \Lambda(x) $ if $ V(Q^{NGI},\phi,\phi^{*}) $ has a non trivial dependence on $ Q^{NGI} $ then equating the variation of the action to zero implies:
\begin{equation}
\phi^{*}(t_{0})\phi(t_{0})=0
\end{equation}
Of course, this means that the theory effectively cancels the non gauge invariant terms when the variational principle is used, so gauge invariance is restored effectively. Also  boundary condition which are gauge invariant are obtained.
%%%%%%%%%%%%%%%%%%%%%%%%%%%%%%%%%%%%%%%%%%%%%%%%%%%%%%%%%%%%%%%%%%%%%%%%%%%
\section{Discussions and Conclusions}
We have studied models where the gauge coupling constants, masses, etc are functions of some conserved charge in the universe. We first considered the standard Dirac action, but where the mass and the electromagnetic coupling constant are a function of the charge in the universe and afterwards extended this to scalar fields. For Dirac field in the flat space formulation, the formalism is not manifestly Lorentz invariant, however Lorentz invariance can be restored by performing a phase transformation of the Dirac field.
\\\\In the case where scalar fields are considered, there is the new feature that an initial condition for the scalar field is derived from the action. In the case of the Higgs field, the initial condition require, that the universe be at the false vacuum state at a certain time slice, which is quite important for inflation scenarios. Also false vacuum branes can be studied in a similar approach.
\\\\One should point out that it appears that not all possible boundary condition allow a formulation in this way, which is probably good, because we would like a theory of the boundary condition to restrict such possibilities.\\
Some "spoiling terms" that is terms that break gauge invariance have been shown that in the end they do not contribute to the functional integral \cite{spoil}. We have seen that "spoiling terms" where non gauge invariant charges are introduced, have as a consequences that they induce boundary condition and these boundary condition imply the vanishing of the spoiling terms, and in the special example choose that the universe sits in the false vacuum in a certain time slice
\subsection*{Next to be explored}
Until now we have used just the $ U(1) $ symmetry of a charged scalar field. Next we will explore generalization of these actions using non Abelian charges which could also lead to the determination of initial condition.
We should study in more detail the Higgs inflation, and how the needed boundary condition are obtained.
We will try to find a way where initial and boundary conditions can be obtained for other kind of fields such as Dirac or vector field.
This development can help us to fix boundary condition in some applications to Bag models in the question of confinement, or even for applications in condensed mater theory where we have Fermions in substances.\\

\subsubsection*{Considering a dynamical surface}
So far we have discussed a given constant time surface which is a space-like surface. Also the possibility of a time-like surface has been discussed.
\\This could represent the dynamical surface of a physical object, such surface should be really dynamical, obeys some equation of motion, for this the theory of dynamical membranes can be of used \cite{yuval}. In the quantum mechanical case we will have some associated wave function, with each surface having a definite amplitude. Such wave function could define then the relevant initial condition of the universe.
Also the spoiling term approach can be used to induce the false vacuum boundary condition in a time like surface as well. 

\subsubsection*{Physical meaning of the boundary inducing terms}
We should explore in more depth the physical meaning of these terms that we have seen induce boundary condition, one idea concerning the "spoiling term" is that we introduce terms defined with a "wrong charge" which made them vanish as we have seen. May be the "wrong charge" term played a role in the early universe but they did not leave any remnant today, but they left us the initial condition that allowed for the inflation in the early universe
%%%%%%%%%%%%%%%%%%%%%%%%%%%%%%%%%%%%%%%%%%%%%%%%%%%%%%%%%%%%%%%%%%%%%%%%%%%%%%%%%%%%%%%%%%%%
%%%%%%%%%%%%%%%%%%%%%%%%%%%%%%%%%%%%%%%%%%%%%%%%%%%%%%%%%%%%%%%%%%%%%%%%%%%%%%%%%%%%%%%%%%%%
%%%%%%%%%%%%%%%%%%%%%%%%%%%%%%%%%%%%%%%%%%%%%%%%%%%%%%%%%%%%%%%%%%%%%%%%%%%%%%%%%%%%%%%%%%%%
%%%%%%%%%%%%%%%%%%%%%%%%%%%%%%%%%%%%%%%%%%%%%%%%%%%%%%%%%%%%%%%%%%%%%%%%%%%%%%%%%%%%%%%%%%%%

%%%%%%%%%%%%%%%%%%%%%%%%%%%%%%%%%%%%%%%%%%%%%%%%%%%%%%%%%%%%%%%%%%%%%%%%%%%%%%%%%%%%%%%%%%%%

%%%%%%%%%%%%%%%%%%%%%%%%%%%%%%%%%%%%%%%%%%%%%%%%%%%%%%%%%%%%%%%%%%%%%%%%%%%%%%%%%%%%%%%%%%%%
\end{document}